\documentclass[12pt]{article}
\usepackage{booktabs}  
\usepackage{amsmath}
\usepackage{amsthm}
\usepackage{amssymb}
\usepackage{mathrsfs}
\usepackage{gensymb}
\usepackage{textcomp}  
\usepackage{xcolor}
\usepackage{lipsum}    
\usepackage{graphicx}
\usepackage{caption}
\usepackage{subcaption}
\usepackage{bm, bbm}
\RequirePackage{diagbox, adjustbox}
\usepackage{natbib}
\usepackage[colorlinks=true,linkcolor=black, citecolor=black, urlcolor=black]{hyperref}
\usepackage{multirow}
\usepackage{multicol}
\usepackage{pdflscape}
\usepackage{afterpage}
\usepackage{everypage}
\usepackage{natbib}
\usepackage{float}
\usepackage{url} 

\begin{document}

\def\spacingset#1{\renewcommand{\baselinestretch}%
{#1}\small\normalsize} \spacingset{1}


\begin{center}
{\Large {\bf Missing Data Imputation and Multilevel Conditional Autoregressive Modeling of Spatial End-Stage Renal Disease Incidence}} 
\end{center}

\vspace{.1in} 

\begin{center}
{\large {\bf Supraja Malladi, Indranil Sahoo, QiQi Lu}} \\
{\large {\it Department of Statistical Sciences and Operations Research,\\
Virginia Commonwealth University}} 
\end{center}

\vspace{.1in}
\baselineskip 18truept

\begin{abstract}

End-stage renal disease has many adverse complications associated with it leading to 20-50\% higher mortality rates in people than those without the disease. This makes it one of the leading causes of death in the United States.  This article analyzes the incidence of end-stage renal disease in 2019 in Florida using a multilevel Conditional Autoregressive model under a Bayesian framework at both the Zip Code Tabulation Area and facility levels. The effects of some social factors and indicators of health on the standardized hospitalization ratio of dialysis facilities are quantified. Additionally, as kidney research studies are posed with a great burden due to missing data, we introduce a novel method to impute missing spatial data using spatial state space modeling. The outcomes of this study offer potentially valuable insights for policymakers aiming to develop strategies that enhance healthcare and service quality for disadvantaged populations.  

\end{abstract}

\noindent%
\vfill

\newpage
\spacingset{1.45} 

\section{Introduction}
\label{Introduction}

End-stage kidney/renal disease (ESKD/ESRD) is a disabling condition that often results in premature death, causing permanent kidney failure. It is the fifth and final stage of the chronic kidney disease (CKD) and requires dialysis or transplantation. There has been a staggering 41.8\% increase in ESRD prevalence from 2000 to 2019 in the United States. According to the USRDS 2020 Annual Data Report, nearly 786,000 people in the United States are living with ESRD with 71\% on dialysis and 29\% with a kidney transplant. This accounted for \$37.3 billion of Medicare expenditures during 2019 \citep{burrows2022reported}. In 2019, heart disease, chronic lower respiratory diseases, diabetes, influenza, and pneumonia were among the 10 leading causes of death and accounted for more than 50\% of all deaths occurring in the United States (Source: National Vital Statistics Reports: From the Centers for Disease Control and Prevention, National Center for Health Statistics, National Vital Statistics System, 2006). Among these, diabetes and heart disease are the most common causes of ESRD in addition to high blood pressure (check for more information on \href{https://www.mayoclinic.org/diseases-conditions/kidney-failure/symptoms-causes/syc-20369048}{Mayo Clinic} and \href{https://www.pennmedicine.org/for-patients-and-visitors/patient-information/conditions-treated-a-to-z/end-stage-kidney-disease#:\sim:text=The\%20most\%20common\%20causes\%20of,conditions\%20can\%20affect\%20your\%20kidneys}{Penn Medicine} websites).

The standardized hospitalization ratio (SHR) of admissions in dialysis facilities is an important indicator of illness in the population. It varies due to various demographic, socio-economic, and behavioral factors. The SHR is calculated numerically as the ratio of the actual number of hospital admissions for the patients in a facility over a specified time period, and the expected number of hospital admissions for the same patients if they were in a facility conforming to the national norm. An SHR equal to 1 indicates that the facility’s overall hospitalization rate equals the national hospitalization rate. The degree to which the facility’s SHR varies from 1 measures the performance of the facility in reducing overall hospitalizations. For example, a facility’s SHR = 1.30 indicates that the facility’s covariate-adjusted
hospitalization rate exceeds the national hospitalization rate by 30\%. Similarly, an SHR = 0.90 indicates that the facility’s hospitalization rate is 10\% below the national hospitalization rates (Technical Notes on the
Standardized Hospitalization Ratio (SHR), July 2021). Since 1995, the Dialysis Facility Reports have utilized hospitalization measures. These reports are instrumental in quality improvement, monitoring, and surveillance efforts conducted by dialysis facilities and ESRD Networks. Hospitalization measures play a crucial role in allocating medical costs and assisting healthcare facilities in delivering cost-effective healthcare services. (Report for the Standardized Hospitalization Ratio, The University of Michigan Kidney Epidemiology and Cost Center, 2016). 

In this study, we analyze the SHR of admissions in 2019 obtained at the facility level, nested within the Zip Code Tabulation Areas (ZCTA) in the state of Florida. We chose Florida as our study region as it had the highest CKD prevalence rate in 2018 among Medicare beneficiaries 65 years and older (Chronic Kidney Disease Disparities in Medicare Fee-For-Service Beneficiaries, Data snapshot August 2020) in addition to being one of the states in the US with the highest prevalence of diagnosed CKD among medicare beneficiaries in 2019 (Kidney Disease Surveillance System, Centers for Disease Control and Prevention). Our goal is to investigate the incidence rate of ESRD in 2019 by modeling associations between the SHR of hospital admissions and some select covariates. The SHR is a good indicator of the quality and effectiveness of care provided, the demand for services, and the ability to prevent re-admissions. It is greatly influenced by factors such as race, gender, excessive alcohol consumption, smoking, income, and social status among others. 

Like most healthcare data, the Medicare certified dialysis facilities data from the Centers for Medicare \& Medicaid Services (CMS)  is also prone to a lot of missing information and this poses a great burden in kidney disease research. Due to various reasons, healthcare facilities often encounter challenges in reporting and recording data, resulting in poor data quality and analysis. The presence of missing data makes it challenging to draw accurate conclusions about different patient groups. Moreover, the need for additional data cleaning and preparation further prolongs the time required for research development \citep{stiglic2019challenges}. In general, it is impossible to address the mechanism that resulted in missing data since no test can prove how missing data is generated \citep{marino2021missing}. \cite{ghazi2022association} examined CKD prevalence and its association with air pollution exposure in Minnesota's Fairview health system, adjusting for variables with more than 5\% missing data using an indicator variable under the assumption of randomness (missing completely at random). \cite{stiglic2019challenges} performed a study on type 2 diabetes mellitus data obtained from three health care centers in Slovenia with a large amount ($\geq 73\%$) of missing data in one of the centers. Simulating missing values from two centers, the study showed a significant model prediction decline with incomplete record removal. 
\cite{blazek2021practical} implemented multivariate imputation by chained equations (MICE) using the `mice' package in R \citep{MICE} on a subset of the National Health and Nutrition Examination Survey (NHANES) 2015-2016 data set (74\% complete cases). The study, focusing on hypertension-kidney disease link in adults, imputed 26 datasets using methods like linear regression, predictive mean matching, and logistic regression via 'mice', observing convergence between actual and imputed data. 
\cite{shara} assessed the performance of listwise deletion, mean of serial measures, adjacent value, multiple imputation, and pattern-mixture in imputing missing renal function data in the Strong Heart Study. The study found that there was no method superior to the rest due to overestimating and underestimating in different methods.

As most of the variables used in this study have missing values and are spatially correlated, we propose a spatial state space modeling approach for missing value imputation. We also make comparisons to imputations generated by the Random Forest method in the `mice' package. Once the missing data is imputed, a multilevel spatial model at both the ZCTA and hospital levels is implemented under a fully Bayesian framework to study the effect of some social factors and indicators of health on the SHR. 

The use of multilevel models to study the incidence of repeated areal Medicare certified dialysis facilities data is notably limited within the existing literature. \cite{Li} proposed a two-level spatiotemporal functional model at the facility level and the region level, to analyze hospitalization rates among dialysis facilities from the USRDS data set. \cite{jamal} conducted a social epidemiology study using survival data and fit four multilevel Poisson regression models to examine the effects of facility and regional variations in the incidence of developing hemodialysis-associated infections (HAIs) as well as its associated factors in more than 6000 patients with ESRD in Japan. Additionally, \cite{SoodE36} evaluated geographic and facility-level variations using a three-level logistic model for data obtained from the Canadian Organ Replacement Register from January 2001 to December 2010. Our dataset comprises multiple hospitals within each ZCTA, effectively yielding replications across spatial units. Hence, we employ a Conditional Autoregressive (CAR) covariance structure to capture the spatial correlation in SHR at the ZCTA level, and an uncorrelated error structure at the hospital level within each ZCTA. This approach enables us to effectively capture and model the spatial dependencies and heterogeneity present in the dialysis facility data. 

The rest of this article is organized as follows. In Section 2, We first present the covariates included in this study along with their missing value information and quality restrictions imposed upon them. Section 3 introduces the spatial local level model and evaluates its performance in spatial missing value imputation. The multilevel spatial CAR model is developed in Section 4 with a modified neighborhood matrix. The details of the model fitting and the results are shown in Section 5. Finally, some concluding remarks close the article in Section 6.

\section{Data Description}\label{analysis}

The Medicare certified dialysis facilities data furnishes comprehensive information about clinical and patient metrics pertaining to Medicare certified ESRD facilities. This dataset encompasses a wide array of data, including patient attributes, treatment trends, hospitalization occurrences, mortality rates, and patterns of transplantation. (see \href{https://data.cms.gov/quality-of-care/medicare-dialysis-facilities}{Medicare dialysis facilities}). For this study, we used the publicly available 2019 CMS Medicare certified dialysis facilities data for the state of Florida. Originally, there were 494 dialysis facilities in Florida across 334 ZCTAs consisting of 27 facilities in Miami, 18 facilities in Jacksonville, 15 in Orlando, 12 in Tampa among others. The SHR of hospital admissions in 2019 was obtained to be modeled as a response variable in Section~\ref{spatial model}. The following variables belonging to the incident population at the facility level for 2019 were obtained as covariates for the study: percentage of patients with diabetes as the primary cause of ESRD, percentage of patients with hypertension as the primary cause of ESRD, percentage of African Americans, number of staff members working for the facility, percentage of septicemia/sepsis cases, and the percentage of female patients.

\begin{figure}[h]
\centering
\includegraphics[width=\textwidth,height=9cm]{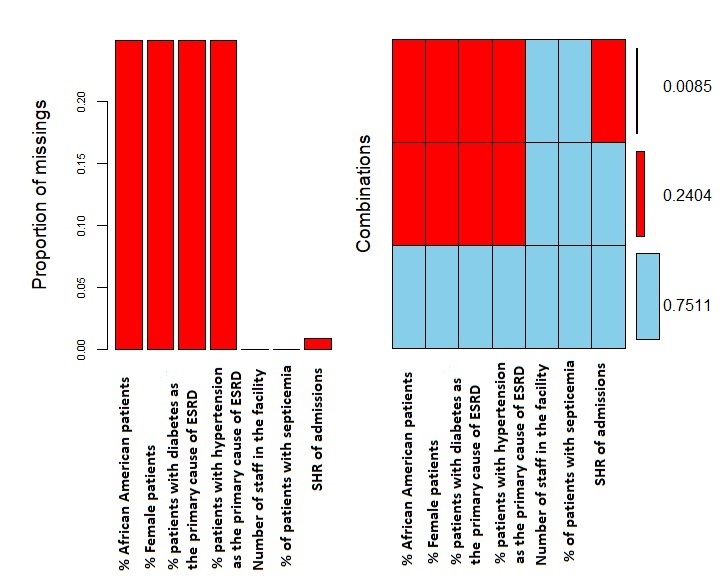}
  \caption{Proportion of missing values in the data (left) and Combinations of missing values in the data (right)}
\label{Fig 1}
\end{figure}

Figure \ref{Fig 1} depicts the proportion of missing values in each of the variables in the data in the left panel and how often each combination of missing values occurs in the right panel. After the variables were selected, the proportions of missing covariates for each hospital were computed and hospitals that had over 80\%  missing covariates were removed from the analysis. This resulted in data for 449 facilities across 325 ZCTAs. Note that there were no missing values for the percentage of patients with septicemia and the number of staff working for the facility. The rest of the covariates had roughly equal proportions of missing values, with 24.04\% missing values for each. In addition, the response variable SHR had 0.85\% of missing values. To incorporate ZCTA-level socioeconomic conditions in our study, we obtained the Federal Poverty Level (FPL) score as an additional covariate from the Social Deprivation Index measures developed using the American Community Survey data at the ZCTA level. The FPL score represents the percent population less than 100\% FPL and indicates the extent of economic disadvantage in a community. It is computed using the formula:
\begin{equation*}
\mbox{FPL score} = \frac{(\mbox{Population} < 0.99~ \mbox{FPL})}{ (\mbox{Total Population})}.
\end{equation*}

\begin{figure}
\centering
\resizebox{\textwidth}{!}{
\begin{tabular}{cc}
  \includegraphics[width=65mm,height=50mm]{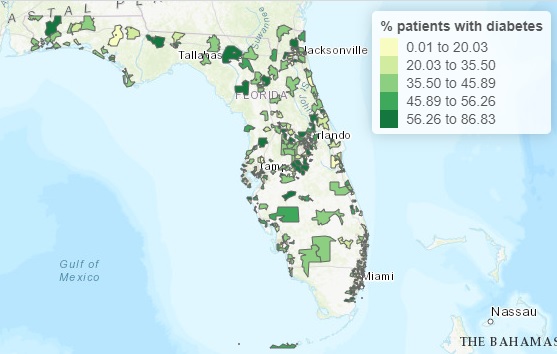} &   \includegraphics[width=65mm,height=50mm]{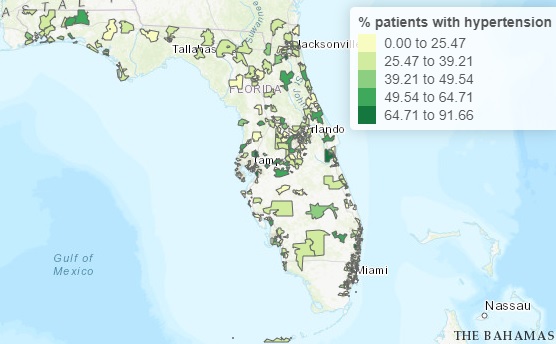} \\
(a)     &          (b)       \\[1pt]
 \includegraphics[width=65mm,height=50mm]{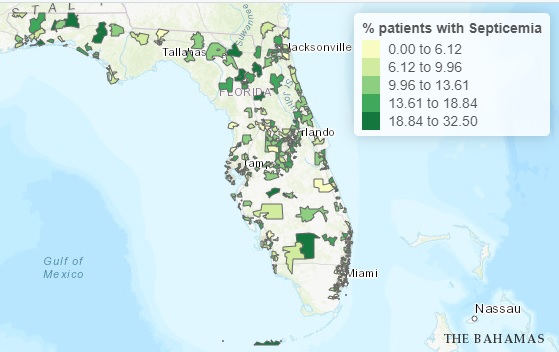} &   \includegraphics[width=65mm,height=50mm]{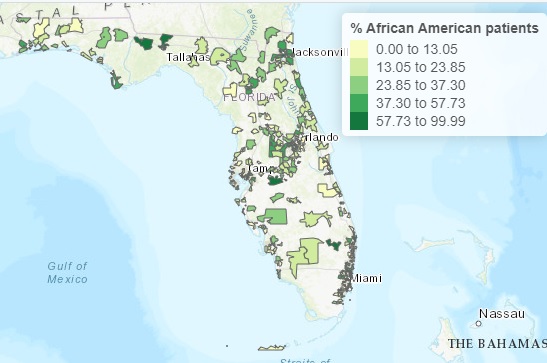} \\
(c)  & (d)  \\[1pt]
 \includegraphics[width=65mm,height=50mm]{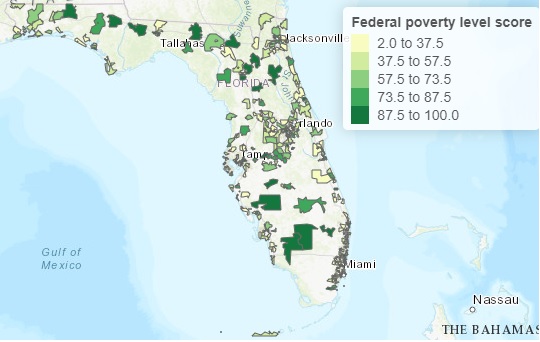} &   \includegraphics[width=65mm,height=50mm]{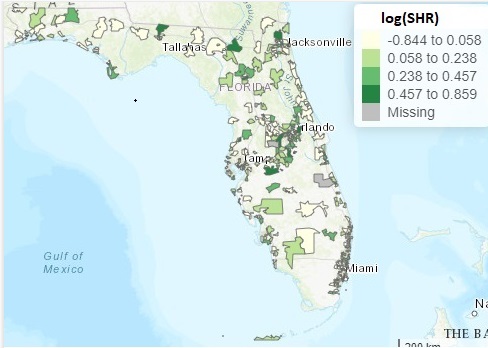} \\
(e)  & (f)  \\ 
\end{tabular}}
\caption{\small Spatial plots of a) Percentage of patients with diabetes, b) Percentage of patients with hypertension as the primary cause of ESRD, c) Percentage of patients with septicemia, d) Percentage of African American patients in the facility, e) Federal poverty level score, f) log(Observed SHR). Average values were computed for ZCTAs with multiple hospitals for each variable.}
\label{Fig 2}
\end{figure}

The spatial distributions of some of the covariates and the SHR are shown in Figure~\ref{Fig 2}. Some of the ZCTAs in Spring Hill, Clermont, Tampa, Orange City, Jacksonville, Orlando, and Gainsville appeared to have high percentages of patients with diabetes and hypertension as the primary causes of ESRD, Federal Poverty Level score, and septicemia. Moreover, Miami-Dade, Palm Beach, Leon, Orange, and Duval counties indicated more than 75\% of African American patients in dialysis facilities in 2019.

\section{Missing value imputation using spatial state space model}\label{method}

We obtained the ZCTAs for each hospital in our data using the End-Stage Renal Disease Quality Incentive Program (ESRD QIP) data provided by the CMS. The distances of each hospital from the centroid of Florida were calculated. Since certain ZCTAs had multiple hospitals, all the facilities belonging to the same ZCTA happened to have the same distance from the centroid. Therefore, to convert the distances to unique values, hospitals were first arranged in the ascending order of distances (including ties) and then insignificant decimal points were added to the end of the duplicated distances. 

Next, we borrowed the idea from state space models in continuous time as described in \cite{koopmansiem}. In continuous time series, the time gaps between consecutive observations are allowed to vary freely unlike in discrete time series where time gaps are equal \citep{Koopman2018}. The local level model in continuous time was slightly adapted by considering the ordered distances as continuous time. As the data has no `time' variable involved, we call our model a `Spatial state space' model, which is described in detail in Section~\ref{ssm}. 

\subsection{Spatial state space model}\label{ssm}

Let $d_{i}$ denote the distance of the 
$i^{th}$ hospital from the centroid of the state for $i = 1,\hdots, M$. Here, $M=449$ denotes the total number of hospitals used in the study. While $d_{i}$ can be any non-negative number, in our case, it is strictly positive. Similar to the sequential ordering of time in continuous time series, distances are arranged in ascending order with $$d_{1} < d_{2} < d_{3} < \hdots < d_{M}.$$

Our spatial local level model for spatial data $\{x_i\}_{i=1}^M$ has the form:
\begin{align}\label{Eq1}
x_{i} = \alpha_{i} + \epsilon_{i}, &\quad \epsilon_{i} \sim N(0, \sigma^2_\epsilon), \\
\alpha_{i+1} =  \alpha_{i} + \eta_{i}, &\quad \eta_{i} \sim N(0, \delta_i \sigma_\eta^2 ),  \label{Eq2}
\end{align}
for $~i = 1,\ldots, M$. Equation (\ref{Eq1}) is referred to as the observation or measurement equation, while Equation (\ref{Eq2}) is called the state or transition equation. Here, $\alpha_i$ is a continuous analog of Gaussian random walk. The measurement errors $\{\epsilon_{i}\}$ are assumed to be independent with a fixed variance $\sigma^2_\epsilon$. Note that the variance of $\epsilon_i$ could be hospital-varying, i.e. $\mbox{Var}(\epsilon_i)=\sigma^2_\epsilon (i)$ for all $i$. The system errors $\{\eta_{i}\}$ are also assumed to be independent. The variance of $\{\eta_{i}\}$ depends on the difference in distance $\delta_i$ between the $i^{th}$ hospital and the $(i-1)^{th}$ hospital after ordering, where $\delta_i = d_i-d_{i-1}$ for $i=1, \ldots, M$ with $d_0=0$. In addition, $\{\epsilon_{i}\}$ and $\{\eta_{i}\}$ are assumed to be independent of each other for all $i$ and independent of the initial state $\alpha_1$.

The parameters in the spatial state space model are estimated via maximum likelihood methods by Kalman filter. Once the model parameters are estimated, the missing values are replaced by their smoothed values with routine applications of the Kalman filter and smoother. It is known that the Kalman filter is powerful in recursive estimation and has good compatibility in handling missing data \citep{saputra}. \cite{adejumo} compared the performances of the Kalman filter algorithm to mean imputation, median imputation, and linear interpolation with 12\% of missing values in time series and found that the Kalman filter algorithm performed better under linear trend structure. For missing dialysis facilities data imputation, a vast amount of dialysis facilities data (22 million data points) obtained in collaboration with the Centers for Dialysis Care and Lytics Health AB was analyzed by \cite{mattias}. The study found that time series models using the Kalman filter gave the best results when compared to traditional imputation methods such as likelihood-based imputation, multiple imputation, and `mice'. Some previous studies which compared different imputation techniques to the Random Forest method in the `mice' package found better imputations using the latter \citep{waljeee,finch,young}. In addition, `mice' is found to impute better with a higher number of trees. While there are many advantages to using `mice', it can be computationally expensive when a higher number of trees are used since they have greater run times \citep{rf}.
 
In order to evaluate the performance of imputations using the spatial state space model by the Kalman filter, we compared our method with the Random Forest method in the `mice' package in Section~\ref{imputingcov}.  

\subsection{Imputing covariates}\label{imputingcov}

The spatial state space model was fit to each of the covariates mentioned in Section~\ref{analysis} individually using the Continuous Time Structural Equation Modelling (ctsem) package in R \citep{ctsempackage} to impute missing values in the data. Since missing values in SHR (response variable) automatically get updated under the Bayesian setting, it was not imputed here.

Additionally, we performed 1000 simulations using both the spatial state space model and the Random Forest method in the `mice' package \citep{MICE} to impute each of the covariates individually for fair comparison although `mice' can perform multivariate imputations. A cross-validation method was used where the complete data were divided into 80\% comprising of the training data set and 20\% comprising of test data. The mean and standard deviation of the symmetric mean absolute percentage error (SMAPE) were calculated to assess the imputations. The SMAPE is computed as follows:

\begin{equation*}
    SMAPE= \frac{100}{N}\sum_{i=1}^{N}\frac{|\hat{x}_i-x_i|}{|\hat{x}_i|+|x_i|},
 \end{equation*}
where $x_i$ is the observed value for the hospital $i$ in the testing dataset, $\hat{x}_i$ is its imputed value, and $N$ is the total number of hospitals in the testing dataset. Table~\ref{Table:1} provides the imputation performance measures of both methods after simulations.  The mean and standard deviation values of SMAPE across both methods are quite close to each other.

\begin{table}[H]
\caption{Performance measures of imputations using state space model and Random Forest in `mice' after 1000 simulations.}
\resizebox{\textwidth}{!}{
\begin{tabular}{lcccc}
\hline
\multirow{2}{*}{\textbf{Variable}} & \multicolumn{2}{c}{\textbf{State space model}}& %
    \multicolumn{2}{c}{\textbf{Random Forest in mice}}\\
\cline{2-5}
 & Mean(SMAPE) & S.D(SMAPE) & Mean(SMAPE) & S.D(SMAPE) \\\hline
\% patients with diabetes as the primary cause of ESRD &0.521&0.048&0.525&0.052\\\hline
\%patients with hypertension as the primary cause of ESRD &0.604&0.052&0.595&0.050\\\hline
\% African American patients&0.783&0.062&0.796&	0.070\\\hline
\% Female patients in the facility&0.250&0.025&0.241&	0.019\\\hline
\end{tabular}}
\label{Table:1}
\end{table}

\begin{figure}
\centering
\begin{tabular}{cc}
  \includegraphics[width=120mm,height=50mm]{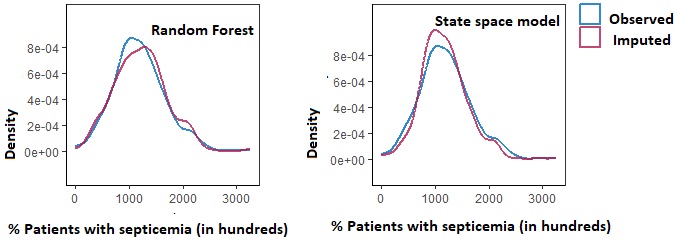}\\
  (a)\\
  \includegraphics[width=120mm,height=50mm]{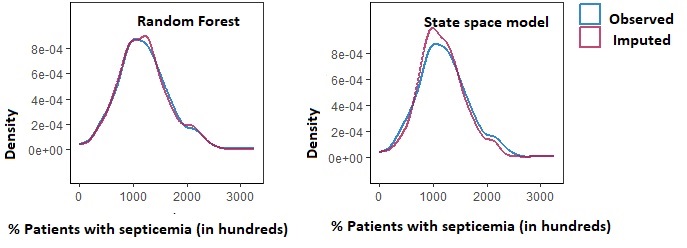}\\
  (b)\\
  \includegraphics[width=120mm,height=50mm]{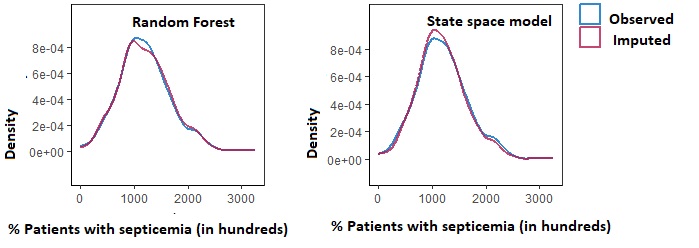}\\
  (c)
\end{tabular}
\caption{Comparing imputations using the Random Forest method in the `mice' package to those using spatial state space model for the percentage of patients with septicemia across (a) 60/40, (b) 70/30, (c) 80/20 train/test split and cross-validation.}
\label{Fig 3}
\end{figure}

\begin{figure}
\centering
\begin{tabular}{cc}
\includegraphics[width=120mm,height=50mm]{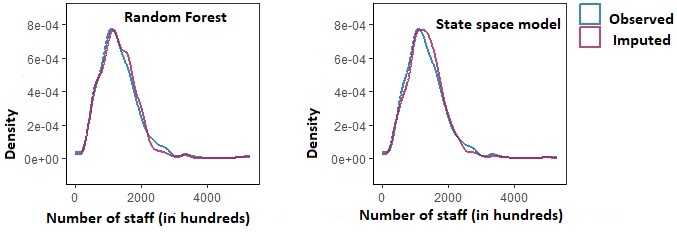} \\
(a)\\
\includegraphics[width=120mm,height=50mm]{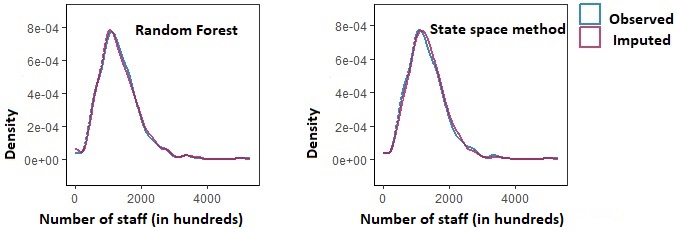}\\
(b)\\
\includegraphics[width=120mm,height=50mm]{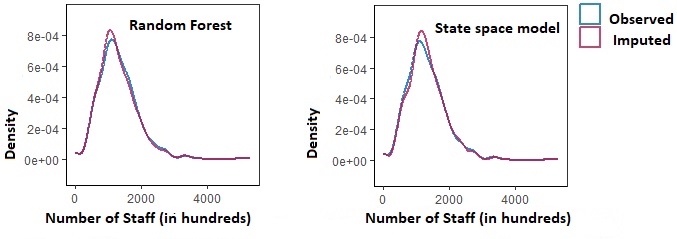} \\
(c)
\end{tabular}
\caption{Comparing imputations using Random Forest method in the `mice' package to those using spatial state space model for the number of staff across (a) 60/40, (b) 70/30, (c) 80/20 train/test split and cross-validation.}
\label{Fig 4}
\end{figure}

To further study the efficacy of the proposed imputation technique, we performed cross-validation on the percentage of sepsis cases and the number of staff since they had no missing values. Data were randomly partitioned into 80\%, 70\%, and 60\% training with 20\%, 30\%, and 40\% testing (missing values) respectively. Imputations using the proposed spatial state space model and Random Forest method in the `mice' package were performed. In all three cases, the imputed distributions of the variables, as shown in Figures \ref{Fig 3} and \ref{Fig 4}, are closer to the true distributions when using our proposed method. 
In addition, the performance measures of both methods are provided in Table \ref{Table:2}. The SMAPE values for the state space model are much lower than those for the Random Forest method, showing the superior imputation performance of the proposed spatial state space model compared to the Random Forest method in `mice'.

\begin{table}[h]
\caption{Performance measures of imputations using state space model and Random Forest in `mice'. }
\resizebox{\textwidth}{!}{
\begin{tabular}{lcccccc}
\hline
\multirow{2}{*}{\textbf{Variable}} & \multicolumn{3}{c}{\textbf{State space model}}& %
    \multicolumn{3}{c}{\textbf{Random Forest in mice}}\\
\cline{2-7}
 & SMAPE(60/40) & SMAPE(70/30) & SMAPE(80/20) & SMAPE(60/40) & SMAPE(70/30) & SMAPE(80/20) \\\hline
\% patients with septicemia &0.162 &0.140&0.093&0.466&0.476&0.443\\\hline
Number of staff working for the facility&0.167 &0.125&0.063&0.485&0.512&0.493 \\\hline
\end{tabular}}
\label{Table:2}
\end{table}

\section{Multilevel Spatial CAR Model}\label{spatial model}

Let $Y_{kj}$ denote the SHR (in the logarithmic scale) of the $j^{th}$ hospital in the $k^{th}$ ZCTA, $j = 1,\hdots, m_k, k = 1,\hdots, K$, where $m_k$ denotes the number of hospitals within the $k^{th}$ ZCTA on which data are available, and $K$ denotes the number of ZCTAs considered. Here, $K=325$.  

The $K \times K$ spatial neighborhood matrix is specified by $\bm{W} = (\omega_{kk'})_{k,k' = 1, \ldots, K}$ in our study. Since not all ZCTAs in Florida have dialysis facilities, we end up with multiple ZCTAs in our dataset having no neighbors. Instead of discarding these ZCTAs with no neighbors, we find their nearest neighbors based on distances among the ZCTAs. For ZCTAs with no neighbors, we designate the ZCTA with the smallest distance from it as its neighbor. Thus, we create an augmented neighborhood matrix $\bm{W}^* = (\omega_{kk'}^*)_{k, k'}$, where $\omega_{kk'}^* = 1$ if ZCTAs  $k$ and $k'$ are adjacent $(k \sim k')$ or if ZCTA $k'$ is closest to ZCTA $k$ when ZCTA $k$ has no adjacent neighbor with dialysis facilities, and is zero otherwise.

Now, we fit log(SHR) in a two-level spatial CAR model introduced in \cite{carbayes}:
\begin{align}
Y_{kj}|\mu_{kj} &\sim N(\mu_{kj} ,\nu^{2}),\\
\mu_{kj} &= \bm{x}^{T}_{kj}\bm{\beta} + O_{kj} + \phi_{k}\label{12}\\
\phi_{k}|\bm{\phi}_{-k} &\sim N \Biggl(\frac{\rho \sum_{k' \neq k}\omega^*_{kk'}\phi_{k'}}{\rho \sum_{k' \neq k}\omega^*_{kk'}+1-\rho}, \frac{\tau^{2}}{\rho \sum_{k' \neq k}\omega^*_{kk'}+1-\rho} \Biggl)\label{13}
\end{align}
where $\bm{\phi} = (\phi_{1},\hdots,\phi_{K})^T$ is the spatial random effect at the ZCTA level and is modeled by a Conditional Autoregressive prior proposed by \cite{leroux2000}.
Also, $\bm{\phi}_{-k} = (\phi_1, \ldots, \phi_{k - 1}, \phi_{k + 1}, \ldots, \phi_K)^T$ denotes the vector of spatial random effects after deleting the random effect corresponding to the $k^{th}$ ZCTA. The logarithms of ZCTA population, denoted by $O_{kj}$, were introduced as the offset for hospital $j$ within ZCTA $k$ in Equation (\ref{12}). $\bm{x}_{kj}$ in Equation (\ref{12}) denotes a vector of all $p$ covariates for hospital $j$ within ZCTA $k$ and $\mu_{kj}$ is the mean function for the hospital $j$ in the ZCTA $k$, which is a linear combination of all the covariates (after accounting for the offset) and the spatial random effect at the ZCTA level.  

In Equation (\ref{13}), $\rho$ is the spatial autocorrelation parameter. $\rho = 0$ corresponds to a lack of spatial interdependence, i.e. the classical multilevel model with 2 levels. On the other hand, $\rho = 1$ leads to the intrinsic CAR model (\cite{congdon2010applied}, p. 183 - 184). Finally, $\nu^{2}$ is the uncorrelated error variance at the hospital level and is an additional scale parameter required when a  Gaussian family is used. 

The following prior distributions were implemented on the model parameters:
\begin{align*}
\bm{\beta} &\sim N(\bm{\mu}_{\beta}, \bm{\Sigma}_{\beta}),\\
\tau^{2} &\sim \mbox{Inverse-Gamma}(a,b),\\
\nu^{2} &\sim \mbox{Inverse-Gamma}(a,b),\\
\rho &\sim \mbox{Uniform}(0,1),
\end{align*}
where $\bm{\mu}_{\beta}$ and $\bm{\Sigma}_{\beta}$ are the mean and variance of the model coefficients, respectively.
For the inverse Gamma priors for the variance components, $\tau^2$ and $\nu^2$, $a$ and $b$ were chosen to be $a = 1, b = 0.01$.

\section{Results}

A fully Bayesian framework has been implemented to fit this model to the data. After discarding 20,000 samples as burn-in, posterior inferences on model parameters were made from 50,000 posterior samples. Posterior predictive distributions were computed for the missing responses, and they were imputed based on 50,000 samples from the posterior predictive distribution. Table~\ref{Table:3} shows the posterior mean and 95\% credible intervals of the model parameters. It is evident that the `percentage of patients with Diabetes as the primary cause of ESRD', the `percentage of patients with Septicemia', and the `ZCTA Federal Poverty Level score' have a significant effect on the SHR of admissions at the facility and ZCTA levels. In addition, the variance components in the model, $\nu^{2}$ and $\tau^{2}$, and spatial autocorrelation $\rho$ are significant. 


The Relative Squared Error (RSE) is used to compare the fitted multilevel spatial model with a persistence (baseline) model, where averages of realized values in ZCTAs were used as predicted values. RSE is a relative metric that divides the squared error of a predictive model by the squared error of a simple model. RSE is not sensitive to the mean and the scale of predictions. It is calculated as follows:
\begin{equation*}
    RSE=\frac{\sum_{k = 1}^K\sum_{j = 1}^{m_k}(Y_{kj}-\widehat{Y}_{kj})^2}{\sum_{k = 1}^K\sum_{j = 1}^{m_k}(Y_{kj}-\Bar{Y}_{k})^{2}}
\end{equation*}
where
\begin{equation*}
\bar{Y}_k= \frac{1}{m_k}\sum_{j = 1}^{m_k} Y_{kj}
\end{equation*}
for $k = 1, \ldots, K$. If the RSE is lower than 1, then the predictive model performs better than the persistence model. The relative squared error between the observed and predicted values of log(SHR) is 0.304, indicating that our model performs well.

\begin{table}[!]
\caption{Model summary of the two-level CAR model}
\resizebox{\textwidth}{!}{
\small
\begin{tabular}{lccc}
\hline
&\textbf{Mean} &\textbf{2.5\%}& \textbf{97.5\%} \\\hline
Intercept & -10.3753 & -10.6752 & -10.0745\\\hline
\% with Diabetes as the primary cause of ESRD& -0.0036 &-0.0069 & -0.0002\\\hline
\% with Hypertension as the primary cause of ESRD&-0.0023 & -0.0054 &  0.0008 \\\hline
\% African American patients in the facility& 0.0003&  -0.0021 & 0.0026\\\hline
Number of Staff working for the facility & -0.0034 & -0.0088 & 0.0021 \\\hline
\% with Septicemia in the facility&  0.0222 & 0.0144 & 0.0300\\\hline
\% Female patients in the facility &0.0003&-0.0030& 0.0036 \\\hline
ZCTA Federal Poverty Level score&0.0028& 0.0004&0.0052 \\\hline
$\nu^2$ & 0.0488 &  0.0376&   0.0630\\\hline
$\tau^2$ & 0.2219& 0.1816 & 0.2695\\\hline
$\rho$ &0.0018  & 0.0001&   0.0066 \\\hline
\end{tabular}}
\label{Table:3}
\end{table}

\begin{figure}
\centering
\resizebox{\textwidth}{!}{
\begin{tabular}{cc}
  \includegraphics[width=70mm, height=70mm]{figures/shrobs.jpg} &   \includegraphics[width=70mm,height=70mm]{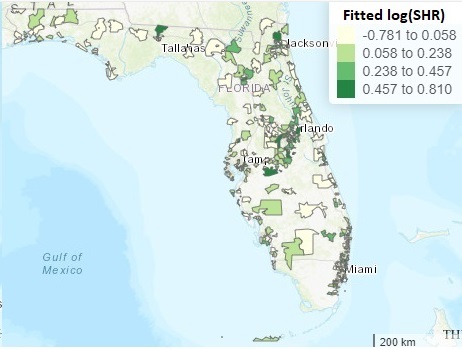} 
\end{tabular}}
\caption{a) Observed log(SHR) and b) fitted log(SHR) across dialysis facilities in Florida. Average values of the SHR were computed for ZCTAs with multiple hospitals before plotting. 
}
\label{Fig 5}
\end{figure}

Figure~\ref{Fig 5} shows the fitted log(SHR) and the observed log(SHR) for comparison across all ZCTAs in Florida containing dialysis facilities. Average values of the log(SHR) were computed for ZCTAs with multiple hospitals before plotting. Some of the ZCTAs in Miami, Oviedo, West Palm-Beach, Orlando, and Bartow indicated high SHR ($>2$) among facilities. While some of the ZCTAs belonging to Naples, Pensacola, Fort Lauderdale, Titusville, and Jacksonville indicated values of the SHR $< 1$. Facilities in Walton, Tallahassee, Gainsville, and Collier have higher predicted values than the observed, while facilities in Lafayette and Hendry have lower values than predicted. Overall, the predicted values fall very close to the observed values of SHR. It is evident that facilities in ZCTAs with high poverty levels, high African American population and higher percentage of patients with septicemia are more susceptible to higher SHR values when compared to the SHR of admissions for the state of Florida. This is consistent with findings from previous studies on nephrology care. The lowest rates of pre-ESRD care among dialysis facilities were discovered in counties with high African American populations and low levls of educational attainment \citep{hua2015,  prakash2010racial}. Moreover,\cite{chang2020} and \cite{ishani2005} indicated an association between septicemia and increased risks of adverse outcomes and hospitalization.

\section{Discussion} \label{Conclusion}

In this study, we implemented a multilevel spatial model to study the effect of some comorbidities, social, and economic factors on the SHR of dialysis facilities in 2019, obtained from the Centers for Medicare and Medicaid Services for the state of Florida. A major challenge in working with such data is that they usually have a lot of missing values. To overcome this challenge, we have proposed a novel spatial state space model for missing data imputation in the Medicare certified dialysis facilities data. We also established that the proposed model is very efficient in imputing missing values based on simulations and comparisons to the Random Forest method in the `mice' package. This method can therefore be used to impute spatial data since it accounts for spatial autocorrelation in the data. Moreover, the multilevel CAR model efficiently incorporated spatial correlation and hierarchy inherent in the data. Thus, studies like ours can be very useful to ESRD networks and dialysis facilities in targeting patient inclusion, improving facility level care, and reducing the number of re-admissions.

In our study,  it was shown that covariates such as ``percentage of patients with diabetes as the primary cause of ESRD", ``percentage of patients with septicemia", and the ``Federal Poverty Level score" had a significant impact on the SHR of admissions when compared to other variables. However, we found facilities in Florida with a higher percentage of patients with diabetes as the primary cause of ESRD to have significantly lower SHR in 2019, which is contrary to what we anticipated. In the past, ESRD burden has been found to be greatly associated with comorbid conditions that include diabetes and hypertension \citep{qi, hsu, mccullough, yang}, poverty at the county level \citep{crews}, being African American \citep{bock}, and deprivation \citep{occelli}. \cite{bilgel2019spatial} studied county level ESRD prevalence averaged over the 2006-2015 period across the US. The study found the strongest ESRD increasing-effects of poverty, high levels of income inequality, and a high percentage of uninsured people in most counties of Florida, Georgia, Carolinas, and Mississippi. The study also found rising hypertension prevalence in Florida in contrast to the results in our study. Facility and ZCTA level ESRD incidence rates are much lower compared to prevalence rates since ESRD is a long-term condition that requires monitoring over a period and incidence only includes one year of data. The risk factors for both prevalence and incidence may also vary due to this. Similar to hypertension, ``the percentage of African Americans" were not significant for Florida, although these factors are well established as important contributors to ESRD in literature. Since we also focus on missing spatial data imputation for a single year, we did not use prevalence rates for our study. 
Although age was not included as a covariate in our study, the state of Florida has been one of the states with the highest percentage of the older population (65 years and above) over the past few years. Therefore, a spatial causal inferential framework is needed to establish the relationship between SHR and the demographic, socio-economic, behavioral, and environmental covariates while accounting for potential confounding variables like age, race, comorbidities, and access to healthcare. Spatial causal inference can help identify spatially varying risk factors, such as environmental exposures, socioeconomic disparities, and healthcare accessibility, that contribute to ESRD prevalence and incidence. Moreover, it enables the evaluation of the effectiveness of interventions and policies targeted at reducing the burden of ESRD within specific geographic areas. By understanding spatial causality, policymakers can allocate resources more efficiently, implement targeted prevention strategies, and tailor healthcare interventions to address the specific needs of high-risk populations in particular regions. \cite{reich2021review} reviews computational methods for causal inference under the spatial setting, which can be easily inherited into kidney research studies.

\bibliographystyle{apalike}

\bibliography{bibfile}
\end{document}